\documentclass{ws-ijmpd}
\usepackage{graphicx}
\usepackage{mathrsfs} 
\renewcommand\draftnote{\phantom{\today}\quad\phantom{\currenttime}\quad WSPC/INSTRUCTION FILE\qquad MZ-TH/07-01}
\begin{document}
\markboth{M.~Reuter and H.~Weyer}
{On the Possibility of Quantum Gravity Effects at Astrophysical Scales}
%
\catchline{}{}{}{}{}
%
\title{ON THE POSSIBILITY OF QUANTUM GRAVITY EFFECTS AT ASTROPHYSICAL SCALES\footnote{Invited contribution to the Int.\ J.\ Mod.\ Phys. D special issue on dark matter and dark energy.}}
\author{MARTIN REUTER}
\author{HOLGER WEYER}
\address{Institute of Physics, University of Mainz\\
Staudingerweg 7 \\
D-55099 Mainz, Germany}
\maketitle
\begin{history}
\received{Day Month Year}
\revised{Day Month Year}
\comby{Managing Editor}
\end{history}
\begin{abstract}
  The nonperturbative renormalization group flow of Quantum Einstein
  Gravity (QEG) is reviewed. It is argued that at large distances
  there could be strong renormalization effects, including a scale
  dependence of Newton's constant, which mimic the presence of dark
  matter at galactic and cosmological scales.
\end{abstract}
\keywords{Quantum gravity; dark matter; galaxies.}
\section{Introduction}

By now it appears increasingly likely that Quantum Einstein Gravity
(QEG), the quantum field theory of gravity whose underlying degrees of
freedom are those of the spacetime metric, can be defined
nonperturbatively as a fundamental, ``asymptotically safe'' theory
\cite{wein}\textsuperscript{-}\cite{max}.
By construction, its bare action is given by a non--Gaussian
renormalization group (RG) fixed point. In the framework of the
``effective average action'' a suitable fixed point is known to exist
within certain approximations. They suggest that the fixed point
should also exist in the exact theory, implying its nonperturbative
renormalizability.

The general picture regarding the RG behavior of QEG as it has emerged
so far points towards a certain analogy between QEG and non--Abelian
Yang--Mills theories, Quantum Chromo Dynamics (QCD) say. For
example, like the Yang--Mills coupling constant, the running Newton
constant $G=G(k)$ is an asymptotically free coupling, it vanishes in
the ultraviolet (UV), i.\,e.\ when the typical momentum scale $k$
becomes large. In QCD the realm of asymptotic freedom is realized for
momenta $k$ larger than the mass scale $\Lambda_{\text{QCD}}$ which is
induced dynamically. In QEG the analogous role is played by the Planck
mass $m_{\text{Pl}}$. It delimits the asymptotic scaling region
towards the infrared (IR). For $k \gg m_{\text{Pl}}$ the RG flow is
well described by its linearization about the non--Gaussian fixed
point. Both in QCD and QEG simple local approximations (truncations)
of the running Wilsonian action (effective average action) are
sufficient above $\Lambda_{\text{QCD}}$ and $m_{\text{Pl}}$,
respectively. However, as the scale $k$ approaches
$\Lambda_{\text{QCD}}$ or $m_{\text{Pl}}$ from above, many
complicated, typically nonlocal terms are generated in the effective
action \cite{gluco}. In fact, in the IR, strong renormalization
effects are to be expected because gauge (diffeomorphism) invariance
leads to a massless excitation, the gluon (graviton), implying
potential IR divergences which the RG flow must cure in a dynamical
way. Because of the complexity of the corresponding flow equations it
is extremely difficult to explore the RG flow of QCD or QEG in the IR,
far below the UV scaling regime, by analytical methods. In QCD,
lattice results and phenomenology suggest that the nonperturbative IR
effects modify the classical Coulomb term by adding a confinement
potential to it which increases (linearly) with distance: $V (r) = -
a/r + \kappa \,r$ \cite{joos}.

The problem of the missing mass or ``dark matter'' is one of the most
puzzling mysteries of modern astrophysics \cite{padman}. It is an
intriguing idea that the apparent mass discrepancy is not due to an
unknown form of matter but rather indicates that we are using the
wrong theory of gravity, Newton's law in the non--relativistic and
General Relativity in the relativistic case.  If one tries to explain
the observed non--Keplerian rotation curves of galaxies or clusters
\cite{combbook} in terms of a modified Newton law, a nonclassical term
needs to be added to the $1/r$-potential whose relative importance
grows with distance \cite{aguirre}. In ``MOND'' \cite{mond}, for
instance, a point mass $M$ produces the potential $\phi (r) = - G M /
r + \sqrt{a_{0} \, G M \,} \, \ln (r)$ and it is tempting to compare
the $\ln(r)$-term to the qualitatively similar confinement potential in
(quenched) QCD. It seems not unreasonable to speculate that the
``confinement'' potential in gravity is a quantum effect which results
from the antiscreening character of quantum gravity \cite{mr} in very
much the same way as this happens in Yang--Mills theory. If so, the
missing mass problem could get resolved in a very elegant manner
without the need of introducing dark matter on an ad hoc basis. In
Refs.\ \refcite{h2,h3} this idea has been explored within a
semi--phenomenological analysis of the effective average action of
quantum gravity \cite{mr}. (See Refs.\ \refcite{bh}-\refcite{h1} for similar work
on gravitational ``RG improvement''. Earlier investigations of IR
quantum gravity effects include Refs.\ \refcite{tsamis}-\refcite{bertoproc}.)

\section{RG running of the gravitational parameters}

The effective average action $\Gamma_{k} [g_{\mu \nu}]$ is a ``coarse
grained'' Wilson type action functional which defines an effective
field theory of gravity at the variable mass scale $k$. Roughly
speaking, the solution to the associated effective Einstein equations
$\delta \Gamma_{k} / \delta g_{\mu \nu} =0$ yields the metric averaged
over a spacetime volume of linear extension $k^{-1}$. In a physical
situation with a typical scale $k$, the equation $\delta \Gamma_{k} /
\delta g_{\mu \nu} =0$ ``knows'' about all quantum effects relevant at
this particular scale. For $k$ fixed, the functional $\Gamma_{k}$
should be visualized as a point in ``theory space'', the space of all
action functionals. When the RG effects are ``switched on'', one
obtains a curve in this space, the RG trajectory, which starts at the
bare action $S \equiv \Gamma_{k \to \infty}$ and ends at the ordinary
effective action $\Gamma \equiv \Gamma_{k \to 0}$
\cite{avact,avactrev,ym}.

The average action is defined in terms of a modified functional
integral over all metrics, $\int \mathcal{D} g_{\mu \nu} \, \exp
\left( - S [g] \right)$, the difference with respect to the
conventional setting being that this integral has a built--in IR
cutoff. It extends only over metric fluctuations with covariant
momenta $p^{2} > k^{2}$.  The modes with $p^{2} < k^{2}$ are given a
momentum dependent $(\text{mass})^{2} \propto \mathcal{R}_{k} \bigl(
p^{2} \bigr)$ and are suppressed therefore. As a result, $\Gamma_{k}$
describes the dynamics of metrics averaged over spacetime volumes of
the size $k^{-1}$, i.\,e.\ $\Gamma_{k} \bigl[ g_{\mu \nu} \bigr]$
gives rise to an effective field theory valid near $k$: when evaluated at
tree level, $\Gamma_{k}$ correctly describes all quantum gravitational
phenomena, including all loop effects, provided the typical
momentum scales involved are all of order $k$. (See Ref.\ \refcite{mr}
and the references therein for a precise definition of these notions.)

The RG trajectory $k \mapsto \Gamma_{k} [\cdot]$ can be obtained
by solving an exact functional RG equation. In practice one has to
resort to approximations. Nonperturbative approximate solutions can be
obtained by truncating the space of action functionals, i.\,e.\ by
projecting the RG flow onto a (typically finite--dimensional) subspace which
encapsulates the essential physics.

The ``Einstein--Hilbert truncation'', for instance, approximates
$\Gamma_{k}$ by a linear combination of the monomials $\int
\!\!\sqrt{g\,} \, R$ and $\int \!\!\sqrt{g\,}$. Their prefactors
contain the running Newton constant $G (k)$ and the running
cosmological constant $\Lambda (k)$. Their $k$-dependence is governed
by a system of two coupled ordinary differential equations.

The flow equations resulting from the Einstein--Hilbert truncation are
most conveniently written down in terms of the dimensionless
``couplings'' $g(k) \equiv k^{d-2} \, G(k)$ and $\lambda(k) \equiv
\Lambda(k) / k^{2}$ where $d$ is the dimensionality of spacetime.
Parameterizing the RG trajectories by the ``RG time'' $t \equiv \ln k$
the coupled system of differential equations for $g$ and $\lambda$
reads $\partial_{t} \lambda = \boldsymbol{\beta}_{\lambda}$,
$\partial_{t} g = \boldsymbol{\beta}_{g}$, where the
$\boldsymbol{\beta}$--functions are given by \cite{mr}
\begin{align}
\begin{split}
\boldsymbol{\beta}_{\lambda}(\lambda, g) 
& = 
-(2-\eta_{\text{N}})\, \lambda + \tfrac{1}{2}\, (4 \pi)^{1-d/2}  \, g \\
& \phantom{{==}}
\times \left[ 2 \, d(d+1) \, \Phi^1_{d/2}(-2\lambda)
- 8 \, d \, \Phi^1_{d/2}(0) 
- d(d+1) \, \eta_{\text{N}} \, \widetilde{\Phi}^1_{d/2}(-2 \lambda) \right]  
\\
\boldsymbol{\beta}_g(\lambda, g) 
& = 
\left(d-2+\eta_{\text{N}} \right) \, g.
\end{split}
\label{10}
\end{align}
Here $\eta_{\text{N}}$, the anomalous dimension of the operator $\int
\!\! \sqrt{g\,} \,R$, has the representation
\begin{align}
\eta_{\text{N}}(g, \lambda) = \frac{g \, B_1(\lambda)}{1-g \, B_2(\lambda) }.
\label{11}
\end{align}
The functions $B_1(\lambda)$ and $B_2(\lambda)$ are defined by
\begin{align}
\begin{split}
B_1(\lambda) & \equiv 
\tfrac{1}{3} \,(4 \pi)^{1-d/2} 
\Bigl[ d(d+1) \, \Phi^1_{d/2-1}(-2\lambda) 
- 6 \, d(d-1) \, \Phi^2_{d/2}(-2\lambda) \Bigr. \\
 & \phantom{{==} \frac{1}{3} \,(4 \pi)^{1-d/2} \Bigl[ \Bigr.} \Bigl.
-4 \, d \,\Phi^1_{d/2-1}(0) - 24 \Phi^2_{d/2}(0) \Big] 
\\
B_2(\lambda) &\equiv
-\tfrac{1}{6} \,(4 \pi)^{1-d/2} \, 
\left[d(d+1) \, \widetilde{\Phi}^1_{d/2-1}(-2\lambda)
-6 \, d(d-1) \, \widetilde{\Phi}^2_{d/2}(-2\lambda) \right].
\end{split}
\label{12}
\end{align}
The above expressions contain the ``threshold functions''
$\Phi^{p}_{n}$ and $\widetilde \Phi^{p}_{n}$. They are given by
\begin{align}
\Phi^p_n(w) &= 
\frac{1}{\Gamma(n)} \int \limits^{\infty}_{0} \!\! \text{d}z~
z^{n-1} \, \frac{ R^{(0)}(z) - z\, R^{(0)\prime}(z)}{\left[ z +  R^{(0)}(z) + w \right]^p\,}
\label{13}
\end{align}
and a similar formula for $\widetilde \Phi^{p}_{n}$ without the
$R^{(0)\prime}$--term. In fact, $R^{(0)}$ is a dimensionless version of
the cutoff function $\mathcal{R}_{k}$, i.\,e.\ $\mathcal{R}_{k} \bigl( p^{2} \bigr)
\propto k^{2} \, R^{(0)} \bigl( p^{2} / k^{2} \bigr)$. Eq.\ \eqref{13}
shows that $\Phi^{p}_{n} (w)$ becomes singular for $w \to -1$. (For
all admissible cutoffs, $z + R^{(0)} (z)$ assumes its minimum value
$1$ at $z=0$ and increases monotonically for $z > 0$.) If $\lambda
>0$, the $\Phi$'s in the $\boldsymbol{\beta}$--functions are evaluated
at negative arguments $w \equiv - 2 \lambda$. As a result, the
$\boldsymbol{\beta}$--functions diverge for $\lambda \nearrow 1/2$ and
the RG equations define a flow on a half--plane only: $- \infty < g <
\infty$, $- \infty < \lambda < 1/2$.

\begin{figure}[t]
\begin{center}
\includegraphics[width=0.95\textwidth]{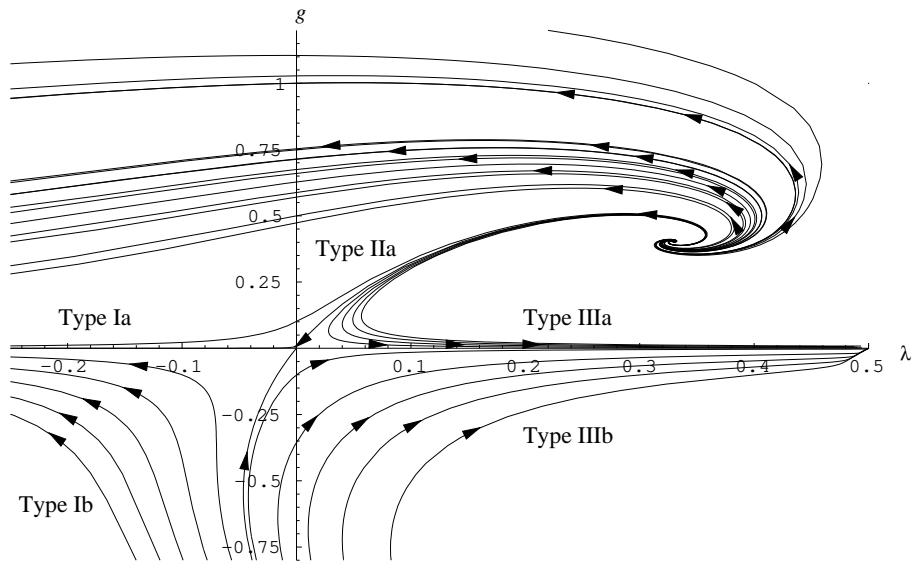}
\end{center}
\caption{RG flow on the $g$-$\lambda$--plane. The arrows point in the direction of decreasing values of $k$.\hfill}{\footnotesize (From Ref.\ \refcite{frank1}.)}
\label{fig1neu}
\end{figure}

This point becomes particularly
clear if one uses a sharp cutoff \cite{frank1}. Then the $\Phi$'s
either display a pole at $w = -1$,
\begin{align}
\Phi^{p}_{n} (w) = 
\frac{1}{\Gamma(n)} \, \frac{1}{p-1} \, 
\frac{1}{(1+w)^{p-1}} \quad \text{for $p>1$},
\label{14}
\end{align}
or, in the special case $p=1$, they have a logarithmic singularity at $w=-1$:
\begin{align}
\Phi^{1}_{n} (w) =
- \Gamma (n)^{-1} \, \ln (1+w) + \varphi_{n}.
\label{15}
\end{align}
The constants $\varphi_{n} \equiv \Phi^{1}_{n} (0)$ parameterize the
residual cutoff scheme dependence which is still present after having
opted for a sharp cutoff. We shall take them equal to the
corresponding $\Phi^{1}_{n} (0)$--value of a smooth exponential cutoff
\footnote{In Fig.\ \ref{fig1neu} the exponential cutoff with ``shape
  parameter'' $s=1$ is used. In $d=4$, the only $\varphi$'s we need
  are $\varphi_{1} = \zeta (2)$ and $\varphi_{2} = 2 \, \zeta (3)$.
  See Ref.\ \refcite{frank1} for a detailed discussion.}, but their
precise value has no influence on the qualitative features of the RG
flow \cite{frank1}. The corresponding $\widetilde \Phi$'s are constant
for the sharp cutoff: $\widetilde \Phi^{1}_{n} (w) = \delta_{p1} /
\Gamma (n+1)$.

From now on we continue the discussion in $d=4$ dimensions. Then, with
the sharp cutoff, the coupled RG equations assume the following form:
\begin{subequations} \label{16}
\begin{align}
\partial_{t} \lambda & =
- \left( 2 - \eta_{\text{N}} \right) \, \lambda
- \frac{g}{\pi} \, \left[ 5 \, \ln (1 - 2 \, \lambda) - \varphi_{2} 
+ \frac{5}{4} \, \eta_{\text{N}} \right]
\label{16a}
\\
\partial_{t} g & = \left( 2 + \eta_{\text{N}} \right) \, g
\label{16b}
\\
\eta_{\text{N}} & =
- \frac{2 \, g}{6 \pi + 5 \, g} \,
\left[ \frac{18}{1 - 2 \, \lambda} 
+ 5 \, \ln (1 - 2 \, \lambda)
- \varphi_{1} + 6 \right].
\label{16c}
\end{align}
\end{subequations}

Solving the system \eqref{16} numerically \cite{frank1} we obtain the
phase portrait shown in Fig.\ \ref{fig1neu}. The RG flow is dominated
by two fixed points $\left( g_{*}, \lambda_{*} \right)$: a Gaussian
fixed point (GFP) at $g_{*} = \lambda_{*} =0$, and a non--Gaussian
fixed point (NGFP) with $g_{*} >0$ and $\lambda_{*} >0$. There are
three classes of trajectories emanating from the NGFP: trajectories of
Type Ia and IIIa run towards negative and positive cosmological
constants, respectively, and the single trajectory of Type IIa
(``separatrix'') hits the GFP for $k\to 0$. The short--distance
properties of QEG are governed by the NGFP; for $k \to \infty$, in
Fig.\ \ref{fig1neu} all RG trajectories on the half--plane $g>0$ run into
this point. \\
\indent 
The conjectured nonperturbative renormalizability of QEG is due to the
NGFP: if it is present in the full RG equations, it can be used to
construct a microscopic quantum theory of gravity by taking the limit
of infinite UV cutoff along one of the trajectories running into the
NGFP, thus being sure that the theory does not develop uncontrolled
singularities at high energies \cite{wein}. By definition, ``QEG'' is the
theory whose bare action $S$ equals the fixed point action $\lim_{k
  \to \infty} \Gamma_{k} \bigl[ g_{\mu \nu} \bigr]$.

Let us pause here for a moment and comment on the physics encoded in
the beta functions \eqref{10} and the flow they imply. They express
the key property of QEG, namely the \textit{antiscreening character of
  the gravitational interaction}. In fact, the scale dependence of
Newton's constant is governed directly by the anomalous dimension
$\eta_{\text{N}}$. In $d=4$ dimensions, say, its flow equation is
$\partial_t g = (2 + \eta_{\text{N}}) \, g$ which translates to
\begin{align} \label{n1}
k \, \tfrac{\partial}{\partial k} \, G (k) & = \eta_{\text{N}} \, G (k)
\end{align}
for the dimensionful $G = g / k^2$. The form of the expression
\eqref{11} for the anomalous dimension illustrates the nonperturbative
character of the beta functions. For $g \, B (\lambda) < 1$, Eq.\ 
\eqref{11} can be expanded as
\begin{align} \label{n2}
\eta_{\text{N}} & = g \, B_1 (\lambda) \sum_{n \geq 0} g^n \, B_2 (\lambda)^n
\end{align}
which shows that even a simple truncation can sum up arbitrarily high
powers of the couplings. It is instructive to consider the
approximation where only the lowest order is retained in \eqref{n2}.
In $d=4$, and for $\lambda (k) \approx 0$, one obtains
$\eta_{\text{N}} = B_1 (0) \, G_0 \, k^2 + \mathcal{O} (G_0^2 \, k^4)$
with $G_0 \equiv G (k=0)$, and integrating \eqref{n1} yields
\begin{align} \label{n3}
G (k) & = G_0 \, \left[ 1 + \tfrac{1}{2} \, B_1 (0) \, G_0 \, k^2
+ \mathcal{O} ( G_0^2 \, k^4 ) \right].
\end{align}
Here $B_1 (0)$ is a $\mathcal{R}_k$-dependent constant which,
however, can be shown to be negative for all admissible cutoff
functions $\mathcal{R}_k$. One sees that at least in the regime where
\eqref{n3} is valid, $G (k)$ is a \textit{decreasing} function of $k$:
Newton's constant is large (small) on low (high) momentum scales.
Interpreting $k$ as an inverse distance, $G$ is an increasing function
of the distance scale. This amounts to the antiscreening behavior
mentioned above.

\begin{figure}[t]
\renewcommand{\baselinestretch}{1}  
\epsfxsize=0.49\textwidth
\begin{center}
\leavevmode
\epsffile{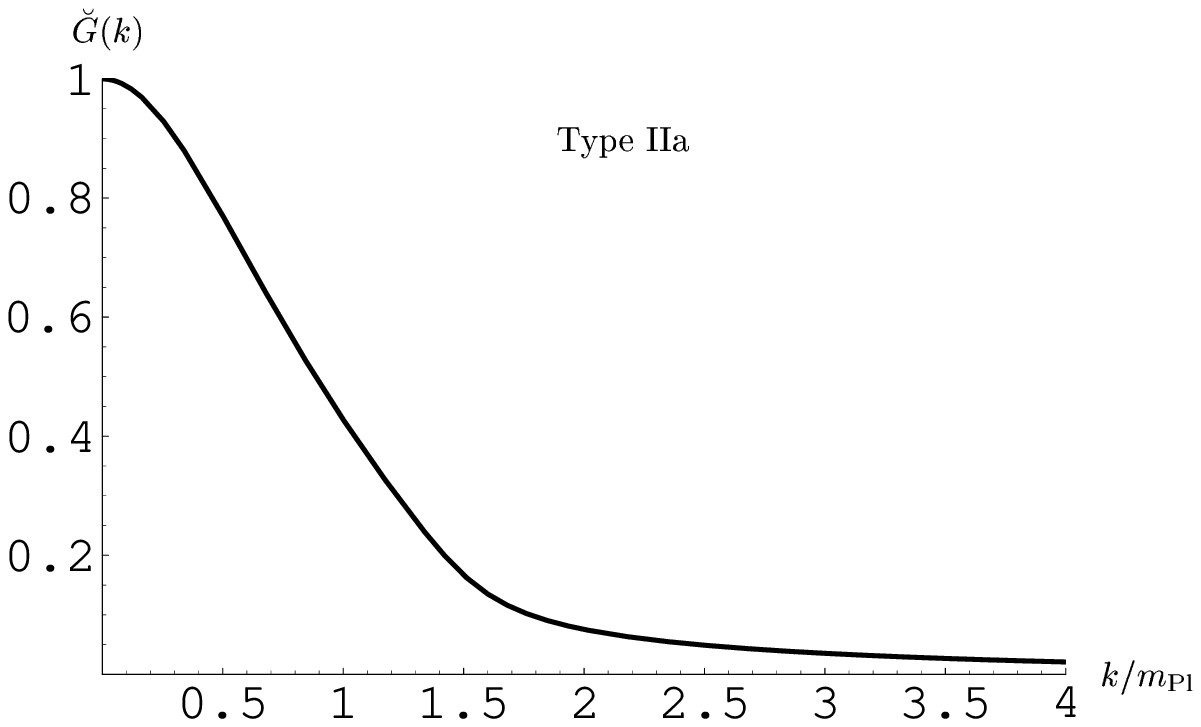}
\epsfxsize=0.49\textwidth
\epsffile{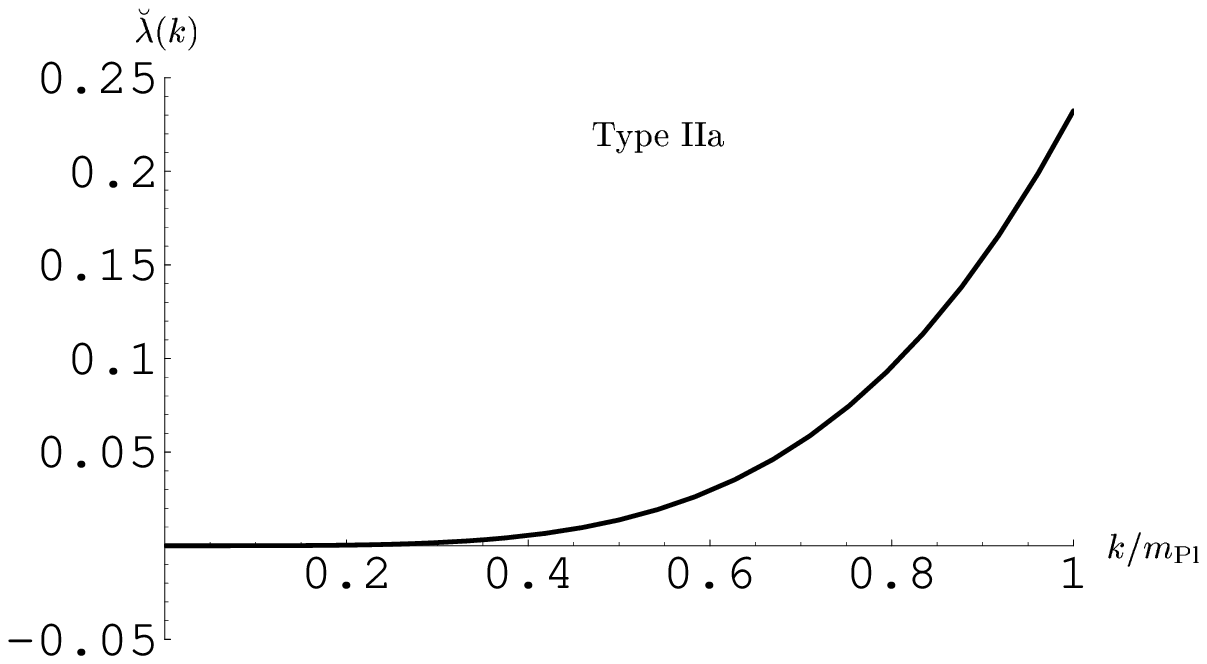}
\end{center}
\vspace*{1.5ex}
\begin{center}
\leavevmode
\epsfxsize=0.49\textwidth 
\epsffile{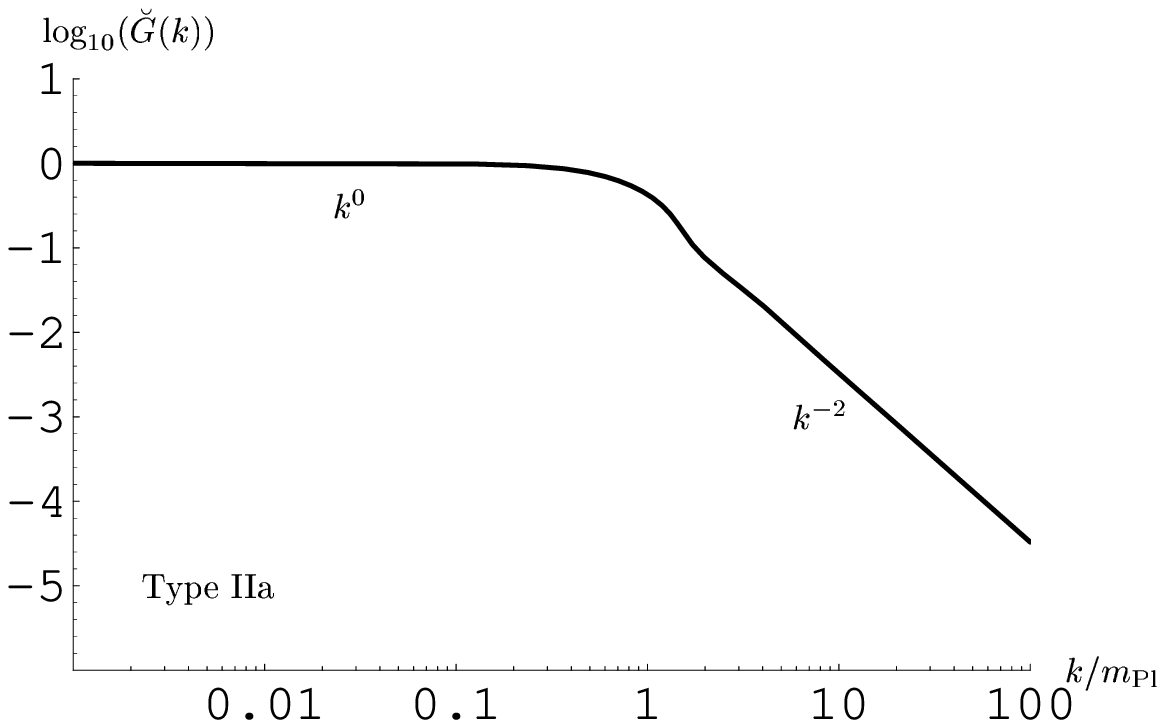}
\epsfxsize=0.49\textwidth
\epsffile{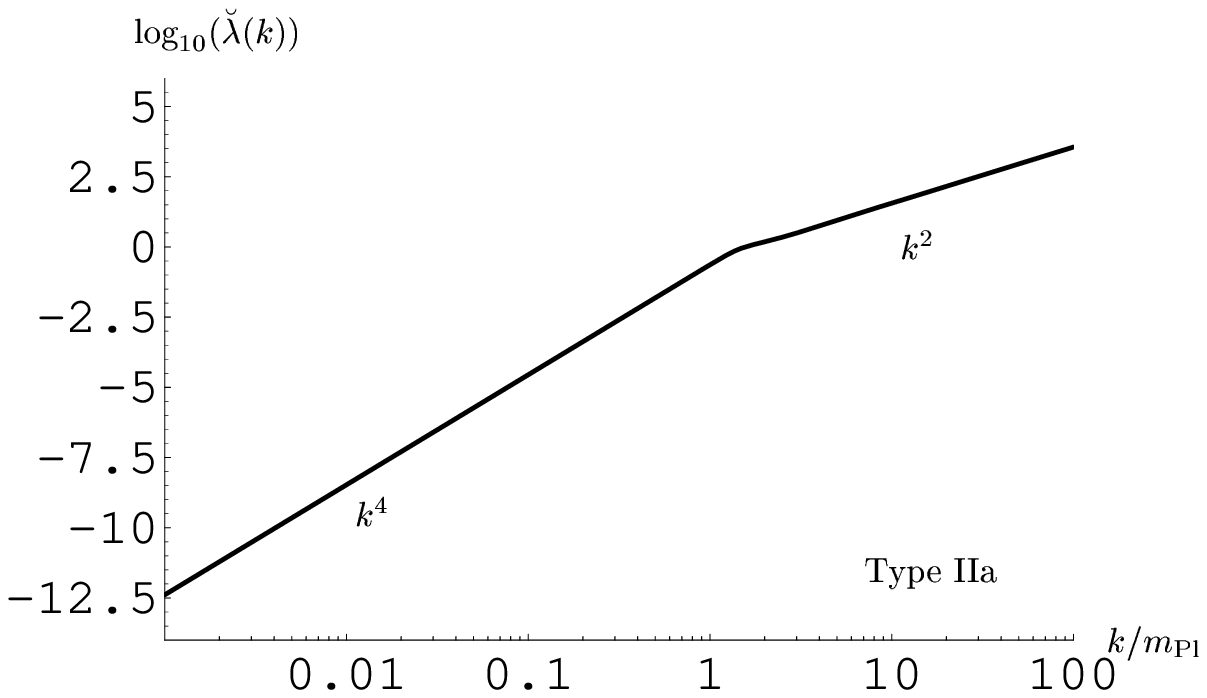}
\end{center}
\vspace*{1.5ex}
\begin{center}
\leavevmode
\epsfxsize=0.49\textwidth 
\epsffile{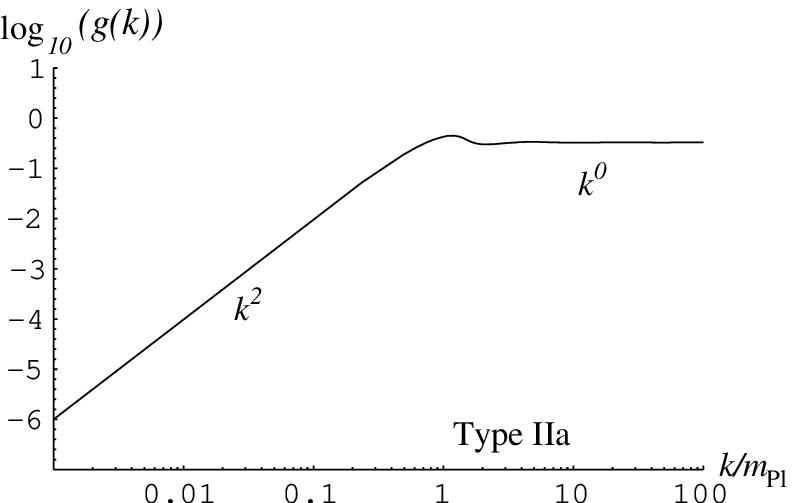}
\epsfxsize=0.49\textwidth
\epsffile{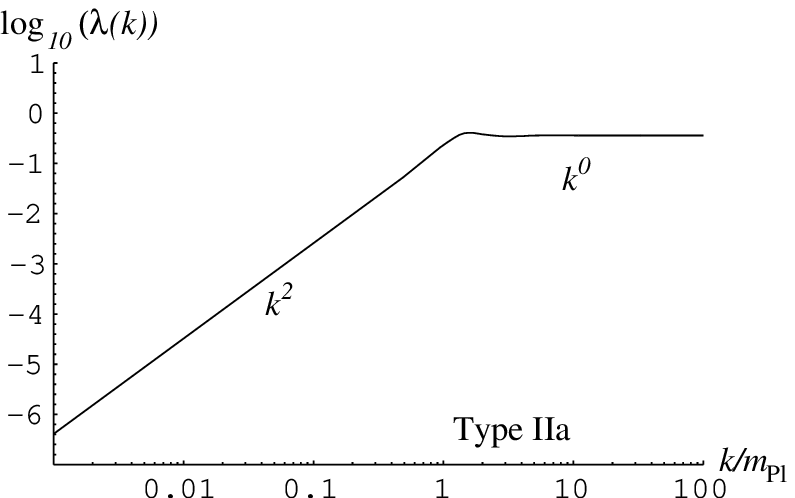}
\end{center}
\caption{Scale dependence of the dimensionful and dimensionless gravitational parameters, respectively. The former are given in units of the Planck mass ($\breve G \equiv G \, m_{\text{Pl}}^2$, $\breve \lambda \equiv \Lambda / m_{\text{Pl}}^2$). The powers of $k$ indicate the scaling laws at the fixed points.}{\footnotesize (From Ref.\ \refcite{frank1}.)}
\label{neu}
\end{figure}

The validity of the approximation \eqref{n3} requires $k \ll
m_{\text{Pl}}$ with the Planck mass defined by the IR value of
Newton's constant, $m_{\text{Pl}} \equiv G_0^{-1/2}$, as well as
$\lambda (k) \approx 0$. As a result, it applies to the lower part of
the separatrix since there both $k / m_{\text{Pl}}$ and $\lambda (k
\to 0)$ are small. In the other regimes numerical methods must be
used. In Fig.\ \ref{neu} we plot both the dimensionful and
dimensionless Newton and cosmological constants along the separatrix as
a function of $k$. One finds that $G (k)$ decreases monotonically with
the momentum scale all the way from $k=0$ up to $k \text{``$=$''}
\infty$. For $k \to \infty$ the scaling governed by the NGFP sets in,
and $G (k) \approx g_\ast / k^2$ vanishes $\propto 1 / k^2$ for $k \to
\infty$. The cosmological constant, on the other hand, increases
monotonically with $k$ and diverges $\propto k^2$ in the NGFP regime.
The logarithmic plots in Fig.\ \ref{neu} illustrate that for most
$k$-values the trajectory is either close to the NGFP or the GFP and
follows the corresponding power law scalings. At $k \approx
m_{\text{Pl}}$ it ``crosses over'' very rapidly from the NGFP to the
GFP. The trajectories of Type Ia have similar properties, the main
difference being that $\Lambda (k)$ becomes negative below a certain
scale.\\
\indent The trajectories of Type IIIa have an important property which
is not resolved in Fig.\ \ref{fig1neu}. Within the Einstein--Hilbert
approximation they cannot be continued all the way down to the
infrared ($k=0$) but rather terminate at a finite scale
$k_{\text{term}} >0$. At this scale they hit the singular boundary
$\lambda = 1/2$ where the $\boldsymbol{\beta}$--functions diverge. As
a result, the flow equations cannot be integrated beyond this point.
The value of $k_{\text{term}}$ depends on the trajectory considered.

\begin{figure}[t]
\includegraphics[width=0.95\textwidth]{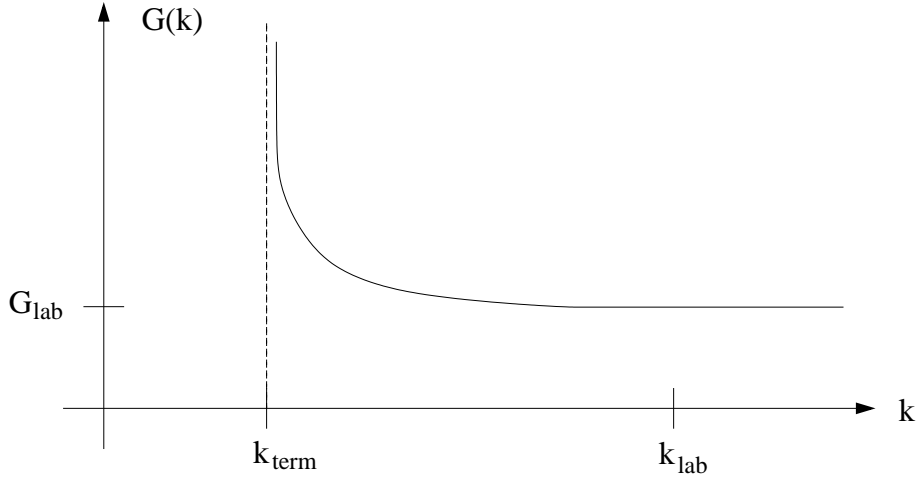}
\caption{Schematic behavior of $G (k)$ for trajectories of
Type IIIa.\hfill}{\footnotesize (From Ref.\ \refcite{h3}.)} 
\label{fig2}
\end{figure}

In Ref.\ \refcite{frank1} the behavior of $g$ and $\lambda$ close to the
boundary was studied in detail. The aspect which is most interesting
for the present discussion is the following. As the trajectory gets
close to the boundary, $\lambda$ approaches $1/2$ from below. In this
domain the anomalous dimension \eqref{16c} is dominated by its pole
term:
\begin{align}
\eta_{\text{N}} \approx - \frac{36 \, g}{6 \pi + 5 \, g} \,
\frac{1}{1 - 2 \, \lambda}.
\label{17}
\end{align}
Obviously $\eta_{\text{N}} \searrow - \infty$ for $\lambda \nearrow
1/2$, and eventually $\eta_{\text{N}} = - \infty$ at the boundary.
This behavior has a dramatic consequence for the (dimensionful) Newton
constant. Since the running of $G (k)$ is given by $\partial_{t} G =
\eta_{\text{N}} \, G$, the large and negative anomalous dimension
causes $G$ to grow very strongly when $k$ approaches $k_{\text{term}}$
from above. This behavior is sketched schematically in Fig.\ 
\ref{fig2}.

At moderately large scales $k$, well below the NGFP regime, $G$ is
approximately constant. As $k$ is lowered towards $k_{\text{term}}$,
$G (k)$ starts growing because of the pole in $\eta_{\text{N}} \propto 1
/ \left( 1 - 2 \, \lambda \right)$, and finally, at
$k=k_{\text{term}}$, it develops a vertical tangent, $\left( \text{d}
  G / \text{d} k \right) \bigl( k_{\text{term}} \bigr) = - \infty$.
The cosmological constant is finite at the termination point: $\Lambda
\bigl( k_{\text{term}} \bigr) = k_{\text{term}}^{2} / 2$.

By fine--tuning the parameters of the trajectory the scale
$k_{\text{term}}$ can be made as small as we like.  Since it happens
only very close to $\lambda = 1/2$, the divergence at
$k_{\text{term}}$ is not visible on the scale of Fig.\ \ref{fig1neu}.
(Note also that $g$ and $G$ are related by a decreasing factor of
$k^{2}$.)

The phenomenon of trajectories which terminate at a finite scale is
not special to gravity, it occurs also in truncated flow equations of
theories which are understood much better. Typically it is a symptom
which indicates that the truncation used becomes insufficient at small
$k$. In QCD, for instance, thanks to asymptotic freedom, simple local
truncations are sufficient in the UV, but a reliable description in
the IR requires many complicated (nonlocal) terms in the truncation
ansatz. Thus the conclusion is that for trajectories of Type IIIa the
Einstein--Hilbert truncation is reliable only well above
$k_{\text{term}}$. It is to be expected, though, that in an improved
truncation those trajectories can be continued to $k=0$. The IR growth
of $G (k)$ can be understood in very general terms \cite{h3} as being
due to an ``instability driven renormalization'' \cite{polinst,oliver0}.

We believe that while the Type IIIa trajectories of the
Einstein--Hilbert truncation become unreliable very close to
$k_{\text{term}}$, their prediction of a growing $G (k)$ for
decreasing $k$ in the IR is actually correct. The function $G (k)$
obtained from the differential equations \eqref{16} should be
reliable, at least at a qualitative level, as long as $\lambda \ll 1$.
For special trajectories the IR growth of $G (k)$ sets in at extremely
small scales $k$ only. Later on we shall argue on the basis of a
gravitational ``RG improvement'' that this IR growth might perhaps be
responsible for the non--Keplerian rotation curves observed in
galaxies.

The other trajectories with $g>0$, the Types Ia and IIa, do not
terminate at a finite scale. The analysis of Ref.\ \refcite{frank1}
suggests that they are reliably described by the Einstein--Hilbert
truncation all the way down to $k=0$.
\section{The RG trajectory ``realized in Nature''}

\begin{figure}[t]
\begin{center}
\includegraphics[width=0.95\textwidth]{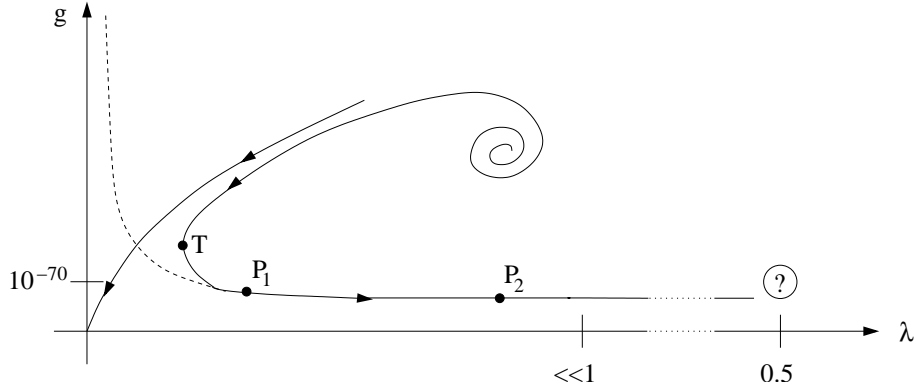}
\end{center}
\caption{Nature's Type IIIa trajectory and the separatrix.
The dashed line is a classical RG trajectory along which $G(k), \Lambda (k) = const$.}{\footnotesize (From Ref.\ \refcite{h3}.)}
\label{fig1}
\end{figure}

In Ref.\ \refcite{h3} we hypothesized that the matter fields present in
the real world do not change the qualitative features of the
Einstein--Hilbert flow and then, on the basis of this hypothesis,
tried to pin down the specific RG trajectory of QEG which is realized
in Nature. Conceptually the procedure is the same as in QED, for
instance, where one fixes the corresponding trajectory by measuring
the electron mass and the fine structure constant. Likewise, in QEG,
the input data are the observed values of Newton's constant and the
cosmological constant. They point towards the highly ``non--generic''
trajectory of Type IIIa sketched in Fig.\ \ref{fig1}.
   
For $k \rightarrow \infty$ it starts infinitesimally close to the
NGFP. Then, lowering $k$, the trajectory spirals about the NGFP and
approaches the ``separatrix'', the distinguished trajectory which ends
at the GFP. It runs almost parallel to the separatrix for a very long
``RG time''; only in the ``very last moment'' before reaching the GFP,
at the turning point T, it gets driven away towards larger values of
$\lambda$. In Fig.\ \ref{fig1} the points P$_1$ and P$_2$ symbolize
the beginning and the end of the regime in which classical general
relativity is valid (``GR regime''). The classical regime starts soon
after the turning point T which is passed at the scale $k_{\rm T}
\approx 10^{-30} m_{\mbox{\scriptsize Pl}}$. In this regime $G (k)$
and $\Lambda (k)$ have almost no $k$-dependence.
   
In Ref.\ \refcite{h3} we speculated that to the right of the point
P$_2$ there starts a regime of strong IR renormalization effects which
might become visible at astrophysical and cosmological length scales.
As we mentioned already, trajectories of Type IIIa terminate at some
$k \neq 0$ near $\lambda = 1/2$ (close to the question mark in Fig.\ 
\ref{fig1}). Before it starts becoming invalid, the Einstein--Hilbert
approximation suggests that $G$ will increase, while $\Lambda$
decreases, as $\lambda \nearrow 1/2$.
   
The Type IIIa trajectory of QEG which Nature has selected is highly
special in the following sense. It is fine--tuned in such a way that it
gets {\it extremely} close to the GFP before ``turning left''. The
coordinates $g_{\rm T}$ and $\lambda_{\rm T}$ of the turning point are
both very small: $g_{\rm T} \approx \lambda_{\rm T} \approx 10^{-60}$. The
coupling $g$ decreases from $g(k) = 10^{-70}$ at a typical terrestrial
length scale of $k^{-1} = 1$ m to $g(k) = 10^{-92}$ at the solar
system scale of $k^{-1} = 1$ AU, and finally reaches $g(k) =
10^{-120}$ when $k$ equals the present Hubble constant $H_0$.
   
In fact, the Hubble parameter $k = H_0$ is approximately the scale
where the Einstein-Hilbert trajectory becomes unreliable. The
observations indicate that today the cosmological constant is of the
order $H_0^2$. Interpreting this value as the running $\Lambda(k)$ at
the scale $k = H_0$ we have $\Lambda (H_0) \approx H_0^2$; as a
result, the dimensionless $\lambda(k)$, at this scale, is of order
unity: $\lambda(H_0) \equiv \Lambda(H_0)/H_0^2 = \mathcal{O} (1)$.
Thus one arrives at a conclusion which is quite remarkable and
intriguing: \textit{the scale at which the IR renormalization effects set in,
  if they exist, is predicted to be the present Hubble scale.}

The ``unnaturalness'' of Nature's gravitational RG trajectory has an
important consequence. Because it gets so extremely close to the GFP
it spends a very long RG time in its vicinity because the
$\boldsymbol{\beta}$-functions are small there. As a result, the termination of the
trajectory at $\lambda = 1/2$ is extremely delayed, by 60 orders of
magnitude, compared to a generic trajectory where this happens for $k$
near the Planck mass. This non--generic feature of the trajectory is a
necessary condition for a long classical regime with $G, \Lambda
\approx const$ to emerge, and any form of classical physics to be
applicable.

It was shown \cite{h3} that for any trajectory which actually does
admit a long classical regime \textit{the cosmological constant in the
  classical regime is automatically small}. In fact, the fine--tuning
behind the ``unnatural'' trajectory Nature has selected is of a much
more general kind than the traditional cosmological constant problem
\cite{coscon}: the primary issue is the emergence of a classical
spacetime; once this is achieved, the extreme smallness of the
observed $\Lambda$ (compared to $m^2_{\text{Pl}}$) comes for free.

Stated differently, if the picture based upon the Einstein--Hilbert
truncation is qualitatively correct \textit{all} quantum theories
(i.\,e.\ QEG based upon \textit{any} of its trajectories) have the
property that if it makes any sense at all to use $S = (16 \pi \,
G)^{-1} \, \int \text{d}^4 x \, \sqrt{-g\,}~\left( R - 2 \Lambda
\right)$ as a classical action, then $\Lambda / m^2_{\text{Pl}} \equiv
\Lambda G$ is guaranteed to be a very small number. QEG seems to
resolve the cosmological constant problem in its original form by
restricting the form of possible classical limits.
   
In principle it should be possible to work out the predictions of the
theory for cosmological scales by an ab initio calculation within QEG.
Unfortunately, because of the enormous technical complexity of the RG
equations, this has not been possible in practice yet. In this
situation one can adopt a phenomenological strategy, however. One
makes an ansatz for the RG trajectory which has the general features
discussed above, derives its consequences, and confronts them with the
observations. In this manner the observational data can be used in
order to learn something about the RG trajectory in the
nonperturbative regime which is inaccessible to an analytic treatment
for the time being. Using this strategy, the cosmological consequences
of a very simple scenario for the $k \to 0$ behavior has been worked
out; the assumption proposed in Refs.\ \refcite{cosmo2,elo} is that the IR effects
lead to the formation of a second NGFP into which the RG trajectory
gets attracted for $k \to 0$.  This hypothesis leads to a
phenomenologically viable late--time cosmology with a variety of
rather attractive features. It predicts an accelerated expansion of
the universe and explains, without any fine--tuning, why the
corresponding matter and vacuum energy densities are approximately
equal.

\section{Galaxy rotation curves}

Given the encouraging results indicating that the IR effects are
possibly ``at work'' in cosmology, by continuity, it seems plausible
to suspect that somewhere between solar system and cosmological scales
they should first become visible. In Refs.\ \refcite{h2,h3} we
therefore investigated the idea that they are responsible for the
observed non--Keplerian galaxy rotation curves. The calculational
scheme used there was a kind of ``RG improvement'', the basic idea
being that upon identifying the scale $k$ with an appropriate
geometric quantity comparatively simple (local) truncations
effectively mimic much more complicated (nonlocal) terms in the
effective action \cite{h1}. Considering spherically symmetric, static
model galaxies, the scale $k$ was taken to be the inverse of the
radial proper distance which boils down to $1 / r$ in leading order.
Since the regime of galactic scales turned out to lie outside the
domain of validity of the Einstein--Hilbert approximation (see below)
the only practical option was to make an ansatz for the RG trajectory
$\big \{ G (k), \Lambda (k), \cdots \big \}$ and to explore its
observable consequences. In particular a relationship between the
$k$-dependence of $G$ and the rotation curve $v (r)$ of the model
galaxy has been derived \cite{h2}.

The idea of the approach proposed in Refs.\ \refcite{h1} and \refcite{h2} is to
start from the classical Einstein--Hilbert action $S_{\text{EH}} =
\int \text{d}^{4} x \, \sqrt{-g\,} \, \mathscr{L}_{\text{EH}}$ with
the Lagrangian $\mathscr{L}_{\text{EH}} = \left( R - 2 \, \Lambda
\right) / \left( 16 \pi \, G \right)$ and to promote $G$ and $\Lambda$
to scalar fields. This leads to the modified Einstein--Hilbert (mEH)
action
\begin{align}
S_{\text{mEH}} [g,G,\Lambda] =
\frac{1}{16 \pi} \, \int \!\! \text{d}^{4} x~
\sqrt{-g\,} \bigg\{
\frac{R}{G (x)} - 2 \, \frac{\Lambda (x)}{G (x)}
\bigg\}.
\label{40}
\end{align}
The resulting theory has certain features in common with Brans--Dicke
theory; the main difference is that $G (x)$ (and $\Lambda (x)$) is a
prescribed ``background field'' rather than a Klein--Gordon scalar as
usual. Upon adding a matter contribution the action \eqref{40}
implies the modified Einstein equation 
\begin{align}
G_{\mu \nu} = - \Lambda (x) \,
g_{\mu \nu} + 8 \pi \, G (x) \, \bigl( T_{\mu \nu} + \Delta T_{\mu
  \nu} \bigr).
\label{41}
\end{align}
Here $\Delta T_{\mu \nu}$ is a new contribution to the
energy--momentum tensor due to the $x$-dependence of $G$:
\begin{align}
\Delta T_{\mu \nu} \equiv \frac{1}{8 \pi} \,
\bigl(
D_{\mu} D_{\nu} - g_{\mu \nu} \, D^{2}
\bigr) \, \frac{1}{G (x)}.
\label{42}
\end{align}
(In Refs.\ \refcite{h1} and \refcite{h2} a further contribution, $\theta_{\mu
  \nu}$, was added to the energy--momentum tensor in order to describe
the 4-momentum of the field $G (x)$. Its form is not completely fixed
by general principles. As it does not affect the Newtonian limit
\cite{h2} we set $\theta_{\mu \nu} \equiv 0$ here.) The field
equation \eqref{41} is mathematically consistent provided $\Lambda
(x)$ and $G (x)$ satisfy a ``consistency condition'' which insures
that the RHS of \eqref{41} has a vanishing covariant divergence.

In Ref.\ \refcite{h2} we analyzed the weak field, slow--motion
approximation of this theory for a time--independent Newton constant $G
= G (\mathbf{x})$ and $\Lambda \equiv 0$. In this (modified) Newtonian
limit the equation of motion for massive test particles has the usual
form, $\ddot {\mathbf{x}} (t) = - \nabla \phi$, but the potential $\phi$
obeys a modified Poisson equation,
\begin{subequations} \label{43}
\begin{align}
\nabla^{2} \phi = 4 \pi \, \overline{G} \, \rho_{\text{eff}}
\label{43a}
\end{align}
with the effective energy density
\begin{align}
\rho_{\text{eff}} = \rho + \bigl( 8 \pi \, \overline{G} \, \bigr)^{-1} \,
\nabla^{2} \mathcal{N}.
\label{43b} 
\end{align}
\end{subequations}
In deriving \eqref{43} it was assumed that $T_{\mu \nu}$ describes
pressureless dust of density $\rho$ and that $G (\mathbf{x})$ does not
differ much from the constant $\overline{G}$. We use the
parameterization
\begin{align}
G (\mathbf{x}) = \overline{G} \, \bigl[ 1 + \mathcal{N} (\mathbf{x}) \bigr]
\label{44}
\end{align}
and assume that $\mathcal{N} (\mathbf{x}) \ll 1$. More precisely, the
assumptions leading to the modified Newtonian limit are that the
potential $\phi$, the function $\mathcal{N}$, and typical (squared)
velocities $\mathbf{v}^{2}$ are much smaller than unity; all terms
linear in these quantities are retained, but higher powers
($\phi^{2}$, $\cdots$) and products of them ($\phi \, \mathcal{N}$,
$\cdots$) are neglected. (In the application to galaxies this is an
excellent approximation.)  Apart from the rest energy density $\rho$
of the ordinary (``baryonic'') matter, the effective energy density
$\rho_{\text{eff}}$ contains the ``vacuum'' contribution
\begin{align}
\bigl( 8 \pi \, \overline{G} \, \bigr)^{-1}
\, \nabla^{2} \mathcal{N} (\mathbf{x}) =
\bigl( 8 \pi \, \overline{G}^{2} \, \bigr)^{-1}
\, \nabla^{2} G (\mathbf{x})
\label{45}
\end{align}
which is entirely due to the position dependence of Newton's constant.
Since it acts as a source for $\phi$ on exactly the same footing as
$\rho$ it mimics the presence of ``dark matter''.

As the density \eqref{45} itself contains a Laplacian $\nabla^{2}$,
all solutions of the Newtonian field equation \eqref{43} have a very
simple structure:
\begin{align}
\phi (\mathbf{x}) = \widehat \phi(\mathbf{x}) + 
\tfrac{1}{2} \, \mathcal{N} (\mathbf{x}).
\label{46}
\end{align}
Here $\widehat \phi$ is the solution to the standard Poisson equation
$\nabla^{2} \widehat \phi = 4 \pi \, \overline{G} \, \rho$ containing
only the ordinary matter density $\rho$. The simplicity and generality
of this result is quite striking.

Up to this point the discussion applies to an arbitrary prescribed
position dependence of Newton's constant, not necessarily related to a
RG trajectory. At least in the case of spherically symmetric systems
the identification of the relevant geometric cutoff is fairly
straightforward, $k \propto 1 / r$, so that we may consider the
function $G (k)$ as the primary input, implying $G (r) \equiv G ( k =
\xi / r)$. Writing again $G \equiv \overline{G} \, \left[ 1 +
  \mathcal{N} \right]$ we assume that $G (k)$ is such that 
$\mathcal{N} \ll 1$. Then, to leading order, the potential for a point
mass reads, according to \eqref{46}:
\begin{align}
\phi (r) = - \frac{\overline{G} \, M}{r}
+ \tfrac{1}{2} \, \mathcal{N} (r). 
\label{47}
\end{align}

Several comments are in order here.

\noindent
\textbf{(a) }The reader might have expected to find a term $-
\overline{G} M \, \mathcal{N} (r) / r$ on the RHS of \eqref{47}
resulting from Newton's potential $\phi_{\text{N}} \equiv -
\overline{G} M / r$ by the ``improvement'' $\overline{G} \to G (r)$.
However, this term $\phi_{\text{N}} \, \mathcal{N}$ is of second order
with respect to the small quantities we are expanding in. In the
envisaged application to galaxies, for example, $\phi_{\text{N}} \,
\mathcal{N}$ is completely negligible compared to the $\tfrac{1}{2} \,
\mathcal{N}$--term in \eqref{47}.

\noindent
\textbf{(b) }According to \eqref{47}, the renormalization effects
generate a nonclassical force (per unit test mass) given by
$-\mathcal{N}^{\prime} (r) / 2$ which adds to the classical $1 /
r^{2}$--term. This force is attractive if $G (r)$ is an increasing
function of $r$ and $G (k)$ a decreasing function of $k$. This is in
accord with the intuitive picture of the antiscreening character of
quantum gravity \cite{mr}: ``Bare'' masses get ``dressed'' by virtual
gravitons whose gravitating energy and momentum cannot be shielded and
lead to an additional gravitational pull on test masses therefore.

\noindent
\textbf{(c) }The solution \eqref{47} is not an approximation artifact.
In Ref.\ \refcite{h2} we constructed exact solutions of the full nonlinear
modified Einstein equations (with $\mathcal{N}$ not necessarily small)
which imply \eqref{47} in their respective Newtonian regime. Those
exact solutions can be interpreted as a ``deformation'' of the
Schwarzschild metric ($M \neq 0$) or the Minkowski metric ($M=0$)
caused by the position dependence of $G$. The solutions related to the
Minkowski metric are particularly noteworthy. They contain no ordinary
matter (no point mass), but describe a curved spacetime, a kind of
gravitational ``soliton'' which owes its existence entirely to the
$\mathbf{x}$--dependence of $G$. At the level of Eq.\ \eqref{47} they
correspond to the $M=0$--potential $\phi = \tfrac{1}{2} \,
\mathcal{N}$ which solves the modified Poisson equation if the
contribution $\propto \nabla^{2} \mathcal{N}$ is the only source term.
In the picture where dark matter is replaced with a running of $G$
this solution corresponds to a \textit{pure dark matter halo} containing no
baryonic matter. The fully relativistic $M=0$--solutions might
be important in the early stages of structure formation \cite{h2}.

Let us make a simple model of a spherically symmetric ``galaxy''. For
an arbitrary density profile $\rho = \rho (r)$ the solution of Eq.\ 
\eqref{43} reads
\begin{align}
\phi (r) = \int \limits_{}^{r} \!\! \text{d} r^{\prime}~
\frac{\overline{G} \, \mathcal{M} (r^{\prime})}{{r^{\prime}}^{2}}
+ \tfrac{1}{2} \, \mathcal{N} (r)
\label{48}
\end{align}
where $\mathcal{M} (r) \equiv 4 \pi \int_{0}^{r}
\text{d} r^{\prime} \,  {r^{\prime}}^{2} \, \rho (r^{\prime})$ is the
mass of the ordinary matter contained in a ball of radius $r$. We are
interested in periodic, circular orbits of test particles in the
potential \eqref{48}. Their velocity is given by $v^{2} (r) = r \,
\phi^{\prime} (r)$ so that we obtain the rotation curve
\begin{align}
v^{2} (r) = \frac{\overline{G} \, \mathcal{M} (r)}{r} 
+ \frac{1}{2} \, r \, \frac{\text{d}}{\text{d} r} \, \mathcal{N} (r).
\label{49}
\end{align}

We identify $\rho$ with the density of the ordinary luminous matter
and model the luminous core of the galaxy by a ball of radius $r_{0}$.
The mass of the ordinary matter contained in the core is $\mathcal{M}
(r_{0}) \equiv \mathcal{M}_{0}$, the ``bare'' total mass of the
galaxy. Since, by assumption, $\rho=0$ and hence $\mathcal{M} (r) =
\mathcal{M}_{0}$ for $r > r_{0}$, the potential outside the core is
$\phi (r) = - \overline{G} \, \mathcal{M}_{0} / r + \mathcal{N} (r) /
2$. We refer to the region $r > r_{0}$ as the ``halo'' of the model
galaxy.

As an example, let us make the scale free power law ansatz $G (k)
\propto k^{-q}$. For $q>0$ Newton's constant increases in the IR. We
assume that this $k$-dependence starts inside the core of the galaxy
(at $r<r_{0}$) so that $G (r) \propto r^{q}$ everywhere in the halo.
For the modified Newtonian limit to be realized, the position
dependence of $G$ must be weak.  Therefore we shall tentatively assume
that the exponent $q$ is very small ($0 < q \ll 1$); applying the
model to real galaxies this will turn out to be the case actually.
Thus, expanding to first order in $q$, $r^{q} = 1 + q \, \ln (r) +
\cdots$, we obtain $G (r) = \overline{G} \, \bigl[ 1 + \mathcal{N} (r)
\bigr]$ with
\begin{align}
\mathcal{N} (r) = q \, \ln (\kappa r)
\label{50}
\end{align}
where $\kappa$ is a constant. In principle the point $\overline{G}$
about which we linearize is arbitrary, but in the present context the
usual laboratory value $G_{\text{lab}}$ is the natural choice. In the
halo, Eq.\ \eqref{50} leads to a logarithmic modification of Newton's
potential
\begin{align}
\phi (r) = - \frac{\overline{G} \, \mathcal{M}_{0}}{r} 
+ \frac{q}{2} \, \ln (\kappa r).
\label{51}
\end{align}
The corresponding rotation curve is
\begin{align}
v^{2} (r) = \frac{\overline{G} \, \mathcal{M}_{0}}{r} 
+ \frac{q}{2}.
\label{52}
\end{align}
Remarkably, at large distances $r \to \infty$ the velocity approaches
a constant $v_{\infty} = \sqrt{q/2\,}$. Obviously the rotation curve
implied by the $k^{-q}$--trajectory does indeed become flat at large
distances --- very much like those we observe in Nature.

Typical measured values of $v_{\infty}$ range from $100$ to
$300~\text{km/sec}$ so that, in units of the speed of light,
$v_{\infty} \approx 10^{-3}$. Thus, ignoring factors of order unity
for a first estimate, we find that the data require an exponent of the
order
\begin{align}
q \approx 10^{-6}.
\label{53}
\end{align}
The smallness of this number justifies the linearization with respect
to $\mathcal{N}$. It also implies that the variation of $G$ inside a
galaxy is extremely small. The relative variation of Newton's constant
from some $r_{1}$ to $r_{2}>r_{1}$ is $\Delta G / G = q \, \ln (r_{2}
/ r_{1})$. As the radial extension of a halo comprises only 2 or 3
orders of magnitude the variation between the inner and the outer
boundary of the halo is of the order $\Delta G / G \approx q$, i.\,e.\ 
Newton's constant changes by one part in a million only.

Including the core region, the complete rotation curve reads
\begin{align}
v^{2} (r) = \frac{\overline{G} \, \mathcal{M} (r)}{r} + \frac{q}{2}.
\label{54}
\end{align}
The $r$-dependence of this velocity is in qualitative agreement with
the observations. For realistic density profiles, $\mathcal{M} (r) /
r$ is an increasing function for $ r < r_{0}$, and it decays as
$\mathcal{M}_{0} / r$ for $r>r_{0}$. As a result, $v^{2} (r)$ rises
steeply at small $r$, then levels off, goes through a maximum at the
boundary of the core, and finally approaches the plateau from above.
Some galaxies indeed show a maximum after the steep rise, but
typically it is not very pronounced, or is not visible at all. The
prediction of \eqref{52} for the characteristic $r$-scale where the
plateau starts is $2 \, \overline{G} \, \mathcal{M}_{0} / q$; at this
radius the classical term $\overline{G} \, \mathcal{M}_{0} / r$ and the
nonclassical one, $q/2$, are exactly equal. With $q = 10^{-6}$ and
$\mathcal{M}_{0} = 10^{11} \, M_{\odot}$ one obtains $9~\text{kpc}$,
which is just the right order of magnitude.

The above $v^{2} (r)$ is identical to the one obtained from standard
Newtonian gravity if one postulates dark matter with a density
$\rho_{\text{DM}} \propto 1 / r^{2}$. We see that if $G (k) \propto
k^{-q}$ with $q \approx 10^{-6}$ no dark matter is needed. The
resulting position dependence of $G$ leads to an effective density
$\rho_{\text{eff}} = \rho + q / \bigl( 8 \pi \, \overline{G} \, r^{2}
\bigr)$ where the $1/r^{2}$--term, well known to be the source of a
logarithmic potential, is present as an automatic consequence of the
RG improved gravitational dynamics.

We consider these results a very encouraging indication
pointing in the direction that quantum gravitational renormalization
effects could perhaps explain the observed non--Keplerian galaxy
rotation curves. If so, the underlying RG trajectory of QEG is
characterized by an almost constant anomalous dimension
$\eta_{\text{N}} = - q \approx - 10^{-6}$ for $k$ in the range of
galactic scales.

Is the Einstein--Hilbert truncation sufficient to search for this
trajectory? Unfortunately the answer is no. According to Eq.\ 
\eqref{16c}, $\eta_{\text{N}}$ is proportional to $g$ which is
extremely tiny in the regime of interest, smaller than its solar
system value $10^{-92}$. In order to achieve a $|\eta_{\text{N}}|$ as
large as $10^{-6}$, the smallness of $g$ must be compensated by large
IR enhancement factors. As a result, $\lambda$ should be extremely
close to $1/2$, in which case the RHS of \eqref{16c} is dominated by
the pole term: $\eta_{\text{N}} \approx - \left( 6 \, g / \pi \right)
\, \left( 1 - 2 \, \lambda\right)^{-1}$. Assuming $g \approx 10^{-92}$
as a rough estimate, a $q$-value of $10^{-6}$ would require $1 - 2 \,
\lambda \approx 10^{-86}$. It is clear that when $1 - 2 \, \lambda$ is
so small the Einstein--Hilbert trajectory is by far too close to its
termination point to be a reliable approximation of the true one.
Moreover, $\eta_{\text{N}} \bigl( g (k), \lambda (k) \bigr)$ is not
approximately $k$-independent in this regime. Thus we must conclude
that an improved truncation will be needed for an investigation of the
conjectured RG behavior at galactic scales.

It is clear that the above model of a galaxy is still quite simplistic
and does not yet reproduce all phenomenological aspects of the mass,
size, and angular momentum dependence of the rotation curves for
different galactic systems. In particular $v_{\infty}$ is a universal
constant here and does not obey the empirical Tully--Fisher relation.
As we explained in Ref.\ \refcite{h2} to which the reader is referred for
further details these limitations are due to the calculational scheme
used here (``cutoff identification'', etc.). Usually this scheme can
provide a first qualitative or semi--quantitative understanding, but
if one wants to go beyond this first approximation, a full fledged
calculation of $\Gamma [g_{\mu \nu}]$ would be necessary which is well
beyond our present technical possibilities.
\section{Conclusion}

The above analysis indicates that if the observed non--Keplerian
rotation curves are due to a renormalization effect, the scale
dependence of Newton's constant should be roughly similar to $G (k)
\propto k^{-q}$. Knowing this, it will be the main challenge for
future work to see whether a corresponding RG trajectory is actually
predicted by the flow equations of QEG. For the time being an ab
initio calculation of this kind, while well--defined conceptually, is
still considerably beyond the state of the art as far as the
technology of practical RG calculations is concerned. In contrast to
phenomenological theories such as MOND it is nevertheless possible to
predict at least the scale on which the IR effects are to be expected.
Given the measured values of $G$ and $\Lambda$, the RG trajectory is
fixed. The ``new physic'' is expected to become visible at scales $k$
for which $\lambda (k)$ gets close to 1/2. For the trajectory
``realized in Nature'' this is the case for $k$ slightly above the
present Hubble parameter $H_0$.

For reliable calculation of the RG trajectory in the IR it might help
to rewrite the nonlocal terms generated during the flow in terms of
local field monomials by introducing extra fields besides the metric.
This is a standard procedure in the Wilsonian approach which often
allows for a simple local description of the effective IR dynamics. It
is tempting to speculate that the resulting local effective field
theory might be related to the generalized gravity theory in Ref.\ 
\refcite{moffat} which includes a Kalb--Ramond field; it is fully
relativistic and explains the galaxy and cluster data with remarkable
precision.

\end{document}